# Multiscale aperture synthesis imager

Ruihai Wang[1,†], Qianhao Zhao[1,†,*], Tianbo Wang[1], Mitchell Modarelli[1], Peter Vouras[2], Zikun Ma[3], Zhixuan Hong[1], Kazunori Hoshino[1], David Brady[4,*], and Guoan Zheng[1,*]

[1]Department of Biomedical Engineering, University of Connecticut, Storrs, CT 06269, USA
[2]United States Department of Defence, Washington, DC 20310, USA
[3]Department of Computer Science and Engineering, University of Connecticut, Storrs, CT 06269, USA
[4]Wyant College of Optical Sciences, University of Arizona, Tucson, Arizona 85721, USA
[†]These authors contributed equally to this work
[*]Email: qianhao.zhao@uconn.edu; djbrady@arizona.edu; guoan.zheng@uconn.edu

**Abstract:** Synthetic aperture imaging has enabled breakthrough observations from radar to astronomy. However, optical implementation remains challenging due to stringent wavefield synchronization requirements among multiple receivers. Here we present the multiscale aperture synthesis imager (MASI), which utilizes parallelism to break complex optical challenges into tractable sub-problems. MASI employs a distributed array of coded sensors that operate independently yet coherently to surpass the diffraction limit of single receiver. It combines the propagated wavefields from individual sensors through a computational phase synchronization scheme, eliminating the need for overlapping measurement regions to establish phase coherence. Light diffraction in MASI naturally expands the imaging field, generating phase-contrast visualizations that are substantially larger than sensor dimensions. Without using lenses, MASI resolves sub-micron features at ultralong working distances and reconstructs 3D shapes over centimeter-scale fields. MASI transforms the intractable optical synchronization problem into a computational one, enabling practical deployment of scalable synthetic aperture systems at optical wavelengths.

Synthetic aperture imaging has revolutionized high-resolution observations across diverse fields, from radar and sonar to radio astronomy. By coherently combining signals from multiple small apertures, it can achieve resolution unattainable by single-aperture receivers. These achievements rely fundamentally on precise timing synchronization between received signals from separated apertures[1]. However, translating synthetic aperture techniques to optical domain presents significant challenges due to the much shorter wavelength of light, which demands stringent sub-wavelength synchronization precision. Conventional optical implementations depend heavily on interferometric methods that establish phase synchronization between apertures[2,3,4,5,6,7,8,9]. There approaches often require intricate optical setups and precise alignment maintenance, limiting their practical scalability outside controlled laboratory settings. Alternative approaches like Fourier ptychography synthesize larger apertures in reciprocal space without direct interferometric measurements[10,11,12,13,14,15,16]. However, this method struggles to reconstruct wavefields exhibiting substantial phase variations or $2\pi$ wraps[17,18], making accurate reconstruction mathematically infeasible[22,23]. Wavefront sensing techniques offer another non-interferometric approach to recover phase[19,20,21,22,23,24,25], yet conventional wavefront sensors measure phase gradients rather than absolute phase values, making them suitable mainly for smoothly varying phase aberrations[20,22]. Crucially, establishing wavefield relationships between separate receivers remains a fundamental barrier to practical systems.

    Here we introduce multiscale aperture synthesis imager (MASI), a new imaging architecture that transforms how synthetic aperture imaging can be implemented at optical wavelengths. MASI builds conceptually on the multiscale approach introduced in gigapixel imaging[26,27,28,29], where complex imaging challenges are broken down into tractable sub-problems through parallelism. While previous multiscale approaches focused primarily on expanding field of view, MASI applies the multiscale paradigm to enhance resolution by coherently synthesizing apertures in real space. To achieve this goal, MASI employs a distinctive measurement-processing scheme where image data is captured at the diffraction plane, while computational synchronization and coherent fusion are performed at the object plane in real space. At the diffraction plane, MASI utilizes a distributive array of coded sensors to acquire lensless diffraction data without reference waves or interferometric measurements. The recovered

wavefields from individual sensors are then digitally padded and propagated to the object plane for coherent synthesis. The key innovation enabling MASI is a computational phase synchronization method, which iteratively tunes the relative phase offsets between sensors to maximize the total energy in the reconstructed object. This process functions analogously to wavefront shaping techniques in adaptive optics[30, 31, 32, 33, 34, 35], where phase elements are optimized to deliver maximum light to target regions. Unlike current approaches that perform synthesis in reciprocal space and require overlapping measurement regions to establish phase coherence, MASI implements alignment, phase synchronization, and coherent fusion entirely in real space, eliminating the need for overlapping measurement regions between separated apertures. This fundamental shift allows distributed sensors to function independently while still contributing to a coherent synthetic aperture that overcomes the diffraction limit of a single receiver. By decoupling physical measurement from computational synthesis, MASI transforms what would be an intractable optical synchronization problem into a manageable computational one, enabling practical, scalable implementations of synthetic aperture imaging at optical wavelengths.

## Results

### Principle of MASI and computational synchronization

Figure 1 illustrates the operating principle and implementation of MASI. In Fig. 1a, we demonstrate how MASI surpasses the diffraction limit of a single sensor by coherently fusing wavefields in real space, without requiring reference waves or overlapping measurement regions between receivers. The process begins with capturing raw intensity patterns using an array of coded sensors positioned at different diffraction planes. Each sensor incorporates a pre-calibrated coded surface that enables robust recovery of complex wavefield information via ptychographic reconstruction. After recovering the wavefields, we computationally pad them and propagate them to the object plane in real space for alignment and synthesis.

An important step enabling MASI's performance is to properly synchronize the wavefields from individual sensors that operate independently without overlapping measurement regions. As shown in the fourth column panels of Fig. 1a, we address this challenge by implementing computational phase synchronization between individual sensors. We designate one of the sensors as a reference and computationally determine the phase offsets for all remaining sensors through an iterative optimization process (Methods). The color blocks in the fourth column panels of Fig. 1a represent these recovered phase offsets, with different color hues indicating different phase values. By maximizing the energy concentration in the reconstructed image, this approach ensures constructive interference between all sensor contributions despite their physical separation, eliminating the need for overlapping measurements or reference waves that constrain conventional techniques. With proper synchronization, the real-space coherent fusion in MASI significantly improves resolution compared to what is achievable with a single sensor in the rightmost panel of Fig. 1a (also refer to Supplementary Figs. S1-S2 and Note 1). In contrast with conventional approaches that perform aperture synthesizing at reciprocal space, MASI operates entirely through real-space alignment, synchronization, and coherent fusion, effectively transforming a distributed array of independent small-aperture sensors into a single large virtual aperture.

An enabling factor for MASI's multiscale strategy is the ability to accurately recover wavefield information using individual sensors. As demonstrated in Supplementary Fig. S3, conventional phase retrieval methods like Fourier ptychography suffer from non-uniform phase transfer function[36] that drops to near-zero values for low spatial frequencies, making them blind to slowly varying phase components such as linear phase ramps and step transitions[18]. In contrast, MASI's individual coded sensor successfully recovers both the step phase transition and the linear phase gradient with only a constant offset from ground truth. This robust performance stems from the coded surface modulation, which converts phase variations -- including low-frequency aberrations -- into detectable intensity variations[21, 43]. For example, a linear phase ramp is converted into a spatial shift of the modulated pattern, while other slowly varying phase variations manifest as distortions in the modulated pattern. Supplementary Fig. S4 illustrates how Fourier ptychography fails when attempting synthetic aperture imaging of complex objects with multiple phase wraps. This demonstrates that without proper phase recovery capabilities at the individual sensor level, conventional methods cannot achieve successful synthetic aperture imaging.

Figure 1b demonstrates MASI's field expansion capability. As the recovered wavefields from individual sensors are digitally padded and propagated back to the object plane, diffraction naturally expands each sensor's field of view beyond its physical dimensions, effectively eliminating gaps in the final reconstruction in the rightmost panel in Fig. 1b (also refer to Supplementary Figs. S5-S6 and Note 1). Figure 1c shows the MASI sensor array, which consists of 9 coded sensors arranged in a grid configuration. This multiscale architecture -- breaking the imaging challenge into parallel, independent sub-problems -- enables each sensor to operate without overlaps with others. During operation, piezoelectric stages introduce sub-pixel shifts (~1-2 μm) to individual sensors for ptychogram acquisition[37], enabling complex wavefield recovery and pixel super-resolution reconstruction[38, 39] from intensity-only diffraction measurements. These shifts are orders of magnitude smaller than the millimeter-scale gaps between adjacent sensors, ensuring completely independent operation that could scale to long-baseline optical imaging, similar to distributed telescope arrays[40] in radio astronomy. In MASI, sensors can be positioned on surfaces at different depths and spatial locations without requiring precise alignment. The design tolerance dramatically simplifies system implementation while maintaining the ability to synthesize a larger virtual aperture. The physical prototype in Fig. 1d demonstrates the system's compact form factor and practical deployment in a reflection configuration, where the sensor array is placed at a 45-degree tilted plane.

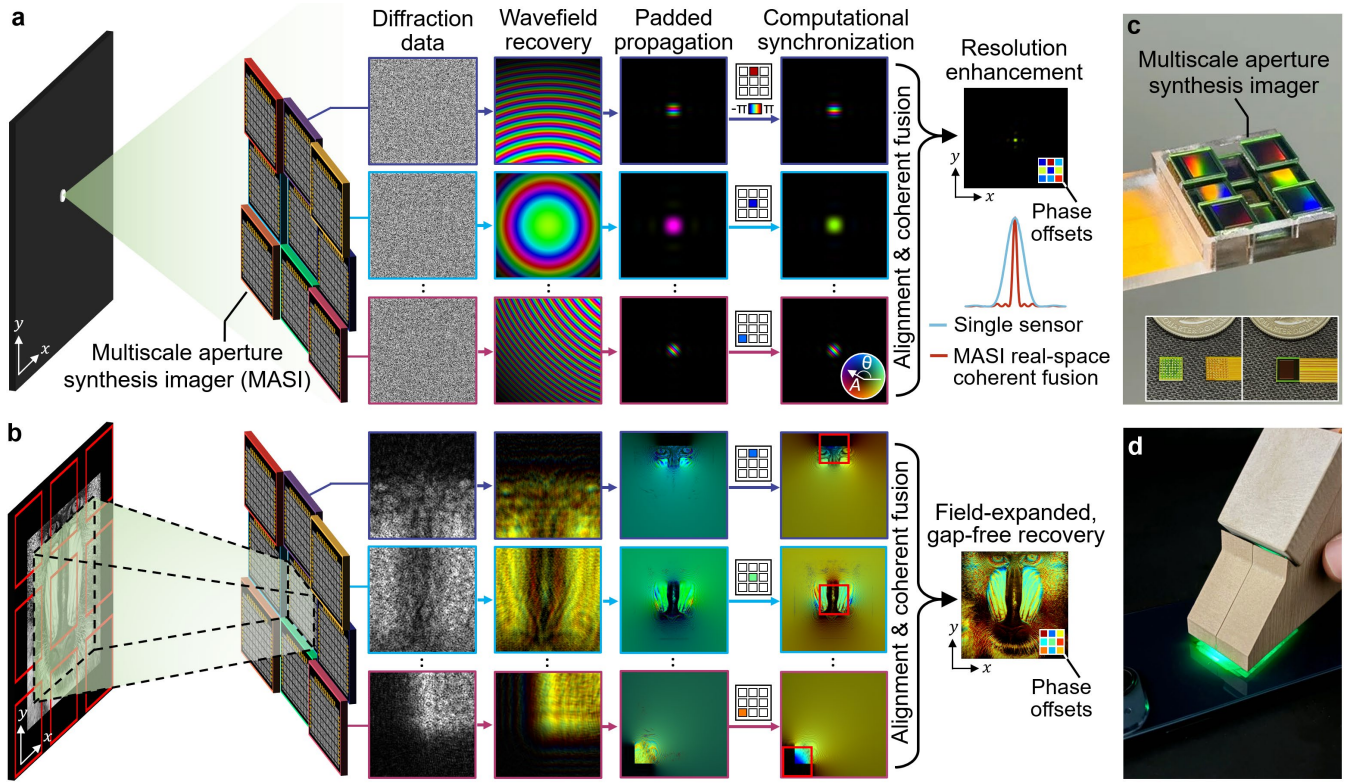

**Fig. 1| Operating principle and implementation of MASI. a**, Resolution enhancement with MASI. Lensless diffraction patterns of a point source are captured by 9 coded sensors (first column). These images are processed to recover the complex wavefields (second column), which are then padded and propagated to the object plane (third column). Through computational phase synchronization (fourth column), MASI synchronizes wavefields from different sensors by optimizing their relative phase offsets to maximize energy in the reconstructed object, without requiring any overlapping measurement regions between individual sensors. In the rightmost panel, the 9 color blocks in the bottom right inset represent the recovered phase offsets of individual sensors, where the phase values are coded with color hues. **b**, Field of view expansion with MASI. As the padded wavefields are propagated from the coded surface plane to the object plane, diffraction naturally expands the field of view beyond individual sensor dimensions, enabling reconstruction despite physical gaps between sensors. **c**, MASI prototype with a compact array of coded sensors. The insets show the coded image sensor and its integration with a customized ribbon flexible cable. **d**, Reflection-mode configuration, where a laser beam illuminates the object surface at ~45 degrees.

The imaging model of MASI can be formulated as follows. We first denote $O(x, y)$ as the object exit wavefield in real space. If the object is a 3D object with a certain non-planar shape, $O(x, y)$ refers to its 2D diffractive field above the 3D object. With the 2D exit wavefield, one can digitally propagate it back to any axial plane and locate the best in-focus position to extract the 3D shape. With $O(x,y)$, the wavefield arriving at the $s^{th}$ coded sensor with a distance $h_s$ can be written as:

$$W_s(x,y) = O(x,y) * psf_{free}(h_s), \quad (1)$$

where $*$ denotes convolution, and $psf_{free}(h_s)$ represents the free-space propagation kernel for a distance $h_s$. Because each sensor is placed at a laterally shifted position $(x_s, y_s)$ and has a finite size, we extract a portion of $W_s(x,y)$ that falls onto the $s^{th}$ coded sensor with $m$ rows and $n$ columns:

$$W_s^{crop}(1:m, 1:n) = W_s(x_s - \frac{m}{2}: x_s + \frac{m}{2}, y_s - \frac{n}{2}: y_s + \frac{n}{2}) \quad (2)$$

The intensity measurement acquired by the $s^{th}$ coded sensor can be written as:

$$I_{s,j}(x - x_j, y - x_j) = |\{W_s^{crop}(x,y) \cdot CS_s(x - x_j, y - x_j)\} * psf_{free}(d)|^2, \quad (3)$$

where $CS_s(x,y)$ represents the coded surface of the $s^{th}$ coded image sensor, and '$* psf_{free}(d)$' models the free-space propagation over a distance $d$ between the coded surface and the sensor's pixel array. Here, the subscript $j$ in $I_{s,j}(x,y)$ represents the $j^{th}$ measurement obtained by introducing a small sub-pixel shift $(x_j, y_j)$ of the coded sensor using an integrated piezo actuator (Methods). Physically, this process encompasses two main steps. The wavefield $W_s^{crop}$ is first modulated by the known coding pattern $CS_s$ upon arriving at the coded surface plane, and then the resulting wavefield propagates a short distance $d$ before reaching the detector.

With a set of acquired intensity diffraction patterns $\{I_{s,j}\}$, the goal of MASI is to recover the high-resolution object wavefield $O(x,y)$ that surpasses the resolution achievable by a single detector. Reconstruction occurs in two main steps. First, the cropped wavefield $W_s^{crop}(x,y)$ is recovered from measurements $\{I_{s,j}\}$ using the ptychographic phase-retrieval algorithm[41, 42]. The recovered wavefield is then padded to its original un-cropped size, forming $W_s^{pad}(x,y)$. Next, each padded wavefield is numerically propagated back to the object plane in real space. The accurate positioning of each coded sensor $(x_s, y_s, h_s)$ is critical for proper alignment and is determined through a one-time calibration experiment (Methods and Supplementary Note 2). Using these calibrated parameters, individual object-plane wavefields are aligned and coherently fused into a single high-resolution reconstruction through computational wavefield synchronization:

$$O_{recover}(x,y) = \sum_s [(e^{i \cdot \varphi_s} \cdot W_s^{pad}(x,y)) * psf_{free}(-h_s)], \quad (4)$$

where $\varphi_s$ is the phase offset for the $s^{th}$ coded sensor. Supplementary Note 3 details our iterative computational phase compensation method that adjusts the unknown $\{\varphi_s\}$ to maximize the integrated energy of the fused reconstruction $O_{recover}$. Supplementary Fig. S7 demonstrates the principle underlying this approach, showing that for objects with both brightfield and darkfield contrast, the total synthesized intensity consistently reaches its maximum when all sensors have their correct phase offsets. This behaviour enables our coordinate descent algorithm in Supplementary Fig. S8, which sequentially optimizes each sensor's phase while maintaining computational efficiency. The effectiveness of this approach is validated in a simulation study in Supplementary Fig. S9, which shows that our method successfully recovers high-fidelity reconstructions from severely distorted initial states, achieving near-zero errors compared to ground truth objects. The proposed computational phase synchronization approach parallels wavefront shaping techniques in adaptive optics[30, 31, 32, 33, 34], ensuring constructive interference and maximum energy recovery. Alternative optimization metrics, such as darkfield minimization or contrast maximization, can also be employed depending on the imaging context. The real-space coherent synchronization in Eq. (4) leads to a resolution enhancement that is unattainable by any single receiver alone, effectively synthesizing a larger effective aperture. By decoupling the phase retrieval and sensor geometry requirements, MASI accommodates flexible sensor placements with minimal alignment constraints, realizing a practical platform for high-resolution, scalable optical synthetic aperture imaging.

**Experimental characterization**
To evaluate and characterize the performance of MASI, we conducted experimental validation under both transmission and reflection configurations. In Fig. 2, we employed a transmission configuration using point-like emitters as test objects. As shown in Fig. 2a, the MASI prototype was positioned to capture diffraction patterns from the single-mode fiber-coupled laser, with piezo actuators enabling small lateral shifts of the sensor array (~1 μm shift per step) to ensure measurement diversity for ptychographic reconstruction. The point-source object served two purposes: it provided validation of resolution improvements through analysis of reconstruction, and it enabled calibration of each sensor's relative position and distance parameters ($x_s, y_s, h_s$). Figure 2b presents the zoomed-in views of the reconstructed complex wavefields from individual sensors of the MASI. The insets of Fig. 2b also show the full fields of view of reconstructions (labeled as 'Full FOV'). These recovered wavefields exhibit distinct fringe patterns corresponding to their respective positions relative to the point object.

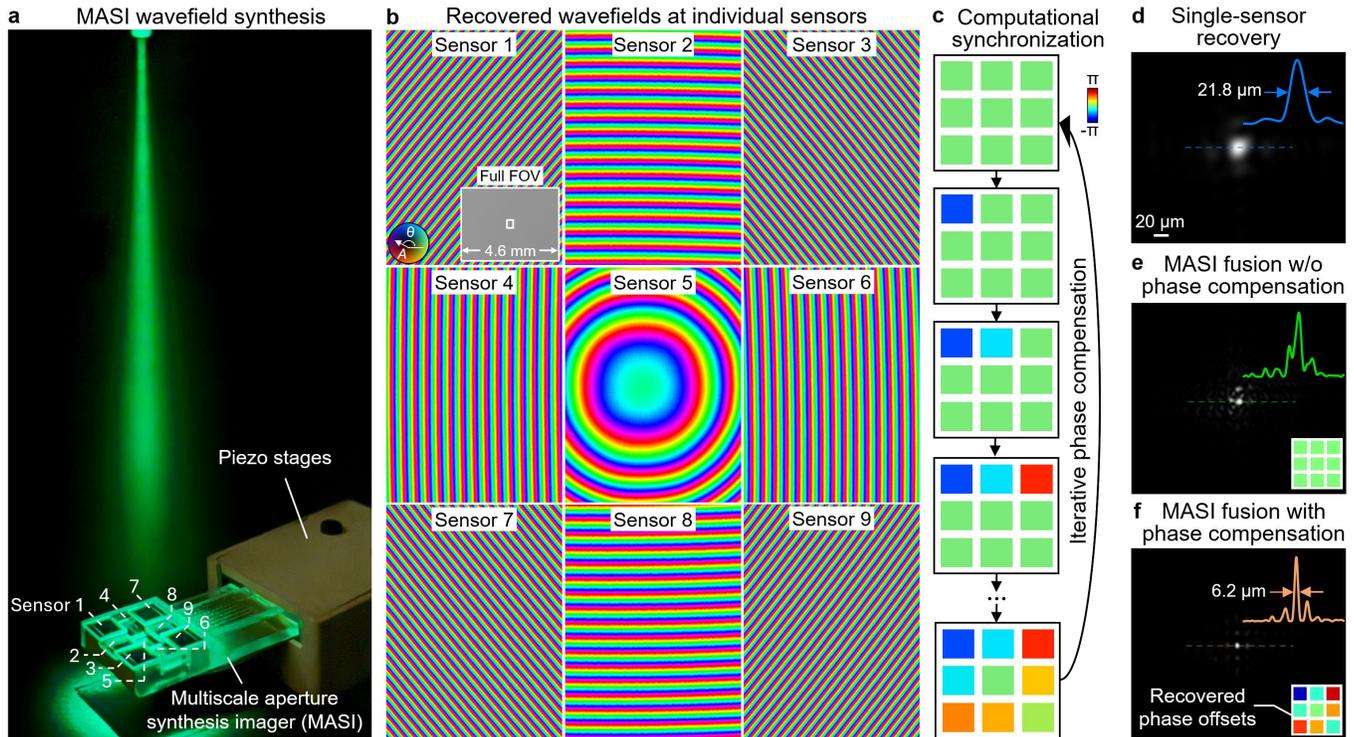

**Fig. 2| Experimental validation of MASI with a point-source object. a**, Schematic of using MASI in a transmission configuration. The numbered coded sensors (1 to 9) are positioned on an integrated piezo stage that introduces controlled sub-pixel shifts. **b**, Complex wavefields recovered from individual sensors of MASI. The color map presents the phase information. The wavefields shown in the main panels are zoomed-in view of the inset panel labelled 'Full FOV'. **c**, The iterative phase compensation process, where phase offsets of individual sensors are digitally turned to maximize the integrated intensity of the object. **d**, The recovered point source using Sensor 5 alone. The limited aperture of a single sensor results in a broadened point source reconstruction. **e**, Result of coherent synthesis without proper phase offset compensation. **f**, MASI coherent synthesis with optimized phase offsets obtained from the computational phase compensation process. The synthesized aperture provides substantially improved resolution. In addition to validating resolution gains, this point-source experiment also provides calibration data ($x_s, y_s, h_s$) for each sensor, enabling precise alignment in subsequent imaging tasks. Supplementary Video 1 visualizes the iterative phase compensation process for computational wavefield synchronization.

To coherently synthesize independent wavefields from different sensors, we implemented the computational phase compensation procedure in Fig. 2c, which iteratively optimizes each sensor's global phase offset to maximize the integrated intensity at the object plane (Supplementary Note 3). The 9 color blocks here represent the recovered phase offsets for the 9 coded sensors, with different color hues indicating different phase values. In the synchronization process, we also employed field-padded propagation to extend the computational window beyond each sensor's physical dimensions for robust real-space alignment. Figures 2d-2f demonstrate the effectiveness of MASI implementation. A single sensor's reconstruction in Fig. 2d shows the point spread function broadening due to its aperture's diffraction limit. Unsynchronized multi-sensor fusion in Fig. 2e yields limited improvement. In contrast, MASI's computational synchronization in Fig. 2f substantially improves resolution. The 9 color blocks in the insets of Figs. 2e-2f show the phase offsets before and after optimization. These results confirm that MASI's computational synchronization effectively extends imaging capabilities beyond individual sensor limitations. Supplementary Video 1 further illuminates the iterative phase synchronization process shown in Fig. 2c.

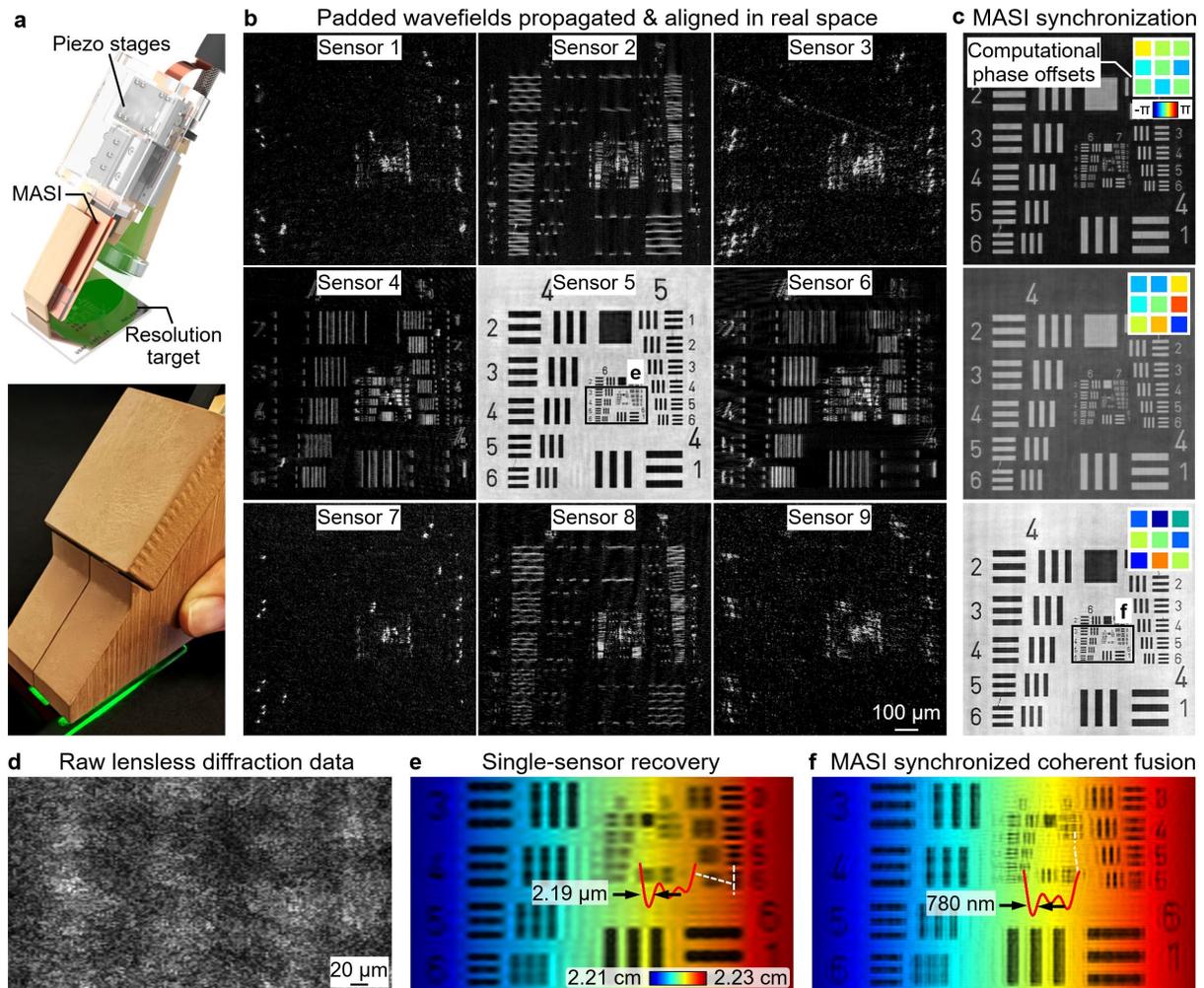

**Fig. 3| MASI high-resolution imaging in reflection configuration with an ultralong working distance of ~2 cm. a**, Schematic of the MASI prototype capturing reflected wavefields from a standard resolution test chart. **b**, Padded wavefields from all nine sensors propagated to the object plane, revealing distinct information captured by each sensor. Each sensor contributes unique high-frequency details of the resolution target, demonstrating the distributed sensing capability of MASI. **c**, Computational phase synchronization process showing different visualization contrasts achieved by varying phase offsets. The 3×3 color blocks in each inset represent the phase values applied to individual sensors, with different hues indicating different phase values. **d**, Raw lensless diffraction data of a zoomed-in region of the resolution target (same region highlighted in the Sensor-5 panel of b). **e**, Single-sensor reconstruction with the raw data in d, resolving linewidths of ~2.19 µm. The color map represents depth information (2.21-2.23 cm) resulting from MASI's tilted configuration relative to the resolution target. **f**, MASI coherent fusion after computational synchronization, resolving 780-nm linewidths at ~2 cm ultralong working distance.

In Fig. 3, we validated MASI in a reflection-mode configuration using a standard resolution test chart positioned at a ~45-degree angle relative to the sensor array. The experimental setup is illustrated in Fig. 3a, where a laser beam illuminates the resolution target and the diffracted wavefield is captured by the MASI. Figure 3b shows the padded, propagated wavefields from all 9 sensors at the object plane, each capturing complementary spatial information of the resolution target. Figure 3c shows the computational phase synchronization process, visualized through different phase offset combinations in the top right insets. One can tune the phase offsets of different sensors to generate additional contrast like the darkfield image in the top panel of Fig. 3c. Comparing the raw diffraction data (Fig. 3d) with single-sensor recovery (Fig. 3e) and synchronized MASI coherent fusion (Fig. 3f) reveals the dramatic resolution enhancement achieved through computational wavefield synthesis. The color variation in Figs. 3e-3f represents depth information ranging from 2.21 cm to 2.23 cm, resulting from MASI's tilted configuration relative to the resolution target. To accurately handle the tilted configuration in this experiment, we implemented a tilt propagation method as demonstrated in Supplementary Fig. S10 (Methods).

As shown in Fig. 3d-3e, quantitative analysis through line traces confirms that MASI resolves features down to 780 nm at a ~2 cm working distance, whereas single-sensor reconstruction is limited to approximately 2.19 μm resolution. This combination of sub-micron resolution and centimeter-scale working distance represents an advancement over conventional imaging approaches, which typically require working distances of one millimeter or less to resolve features at this scale. In Supplementary Fig. S11, we also show different sensor configurations and the corresponding wavefield reconstructions, demonstrating how different sensor combinations affect the resolution. Particularly notable is how the coherent fusion of sensors arranged in complementary positions provides directional resolution improvements along the corresponding spatial dimensions, enabling tailored resolution enhancement for specific imaging applications.

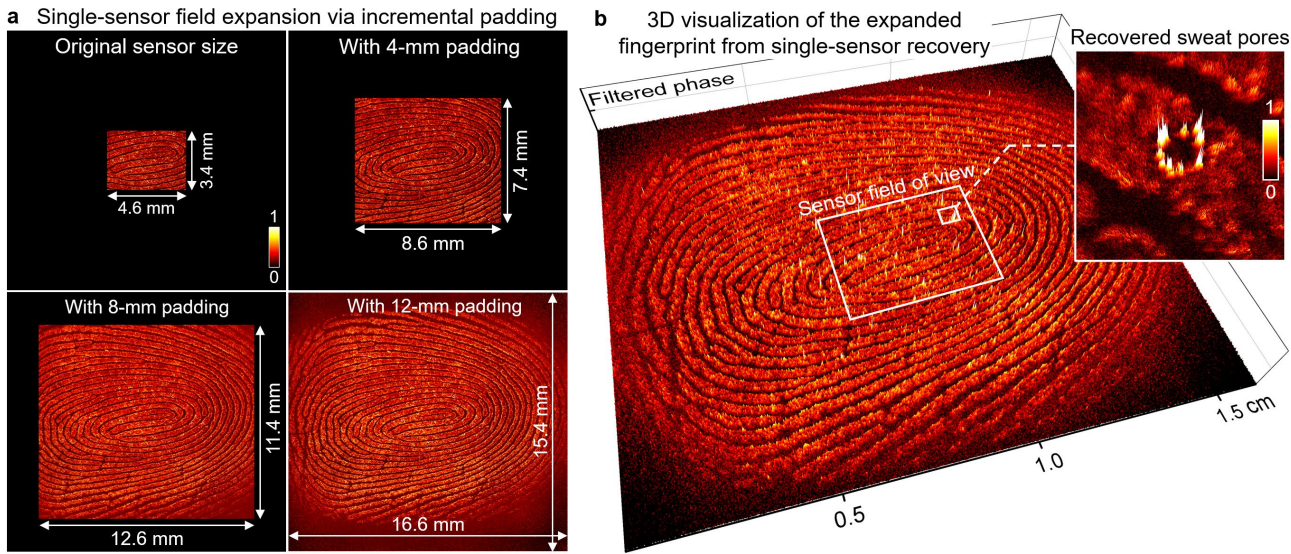

**Fig. 4| Computational field expansion for phase-contrast imaging of fingerprint, achieving 16-fold increase in imaging area compared to the size of the employed sensor. a**, To expand the imaging field of view in MASI, we computationally pad the recovered wavefield at the diffraction plane and propagate it to the object plane. With different padding areas, the reconstruction region grows from the original 4.6 mm × 3.4 mm sensor size to 16.6 mm × 15.4 mm, allowing a much broader view of the fingerprint to be revealed without additional data acquisition. **b**, 3D visualization of the fingerprint across the expanded field, highlighting the detailed fingerprint ridges and the location of individual sweat pores. Supplementary Video 2 visualizes field expansion process via incremental padding.

**Computational field expansion**
Unlike conventional imaging systems, which are often constrained by lens apertures or sensor sizes, MASI leverages light diffraction to recover information at regions outside the nominal detector footprint. This field expansion can be understood as a reciprocal process of wavefield sensing on the detector. When our coded sensor captures diffracted light, it recovers wavefronts arriving from a range of angles, each carrying information about different regions of the object. These angular components contain spatial information extending beyond the physical sensor boundaries. During computational reconstruction, we perform the conjugate operation by back-propagating these captured angular components to the object plane. By padding the recovered wavefield at the detector plane before back-propagation, we effectively allow these angular components to retrace their propagation paths to regions outside the sensor's direct field of view. This process is fundamentally governed by the properties of wave propagation -- the same waves that carry information from extended object regions to our small detector can be computationally traced back to reveal those extended regions. The field expansion arises naturally as the padded recovered wavefields are numerically propagated from the diffraction plane to the object plane (Fig. 1b), effectively reconstructing parts of the object not directly above the sensor.

We also note that both the original and extended wavefield regions maintain the same spatial frequency bandwidth. For the wavefield directly above the coded sensor, the recovered spatial frequency spectrum is cantered at baseband including the zero-order component. As we computationally extend to regions beyond the sensor through padding and propagation, the bandwidth remains constant but shifts to different spatial frequencies based on the angular relationship between the extended location and sensor position. This principle is demonstrated in Fig. 3, where Sensor 5 captures baseband frequencies for the resolution target directly above it, while peripheral sensors capture high-frequency bands of the same region through their extended fields. This distributed frequency sampling across multiple sensors is precisely what enables super-resolution in MASI.

Figure 4a demonstrates the field expansion capability when imaging a fingerprint using a single sensor in MASI. Using 532 nm laser illumination at a 19.5 mm working distance, a single 4.6 mm × 3.4 mm sensor captures only a small portion of the fingerprint's diffraction pattern. By padding the reconstructed complex wavefield and propagating it to the object plane, we can expand the imaging area from the original sensor size of 4.6 mm × 3.4 mm to 16.6 mm × 15.4 mm (with 12-mm padding). To enhance visualization of surface features across this expanded field, the recovered phase maps undergo processing of background subtraction as illustrated in Supplementary Fig. S11a, which improves the contrast of the fine features. Figure 4b shows a 3D visualization of the expanded fingerprint phase map covering an area of 16.6 mm × 15.4 mm, with the inset highlighting resolved sweat pores. Supplementary Fig. S12 further demonstrates the versatility of this approach across different materials including plastic, wood, and polymer surfaces, each revealing distinct micro-textural details. This capability demonstrates how MASI's computational approach transforms a small physical detector into a much larger virtual imaging system with enhanced phase contrast visualization.

Figure 5 further validates MASI's capability for high-resolution, large-area phase-contrast imaging of a mouse brain section. Figure 5a displays the recovered complex wavefields from all 9 individual sensors in the MASI device, each capturing a 4.6 mm × 3.4 mm region. Figure 5b demonstrates the field-expanded recovery using just a single sensor (Sensor 3), where computational padding and propagation transform a small 4.6 mm × 3.4 mm sensor area into a comprehensive 17.0 mm × 13.4 mm phase-contrast visualization of the entire brain section. The recovered field of view is much larger than the physical sensor area highlighted by the green box in Fig. 5b. The speckle-like features visible outside the brain tissue arise from air bubbles in the mounting medium, a sample preparation artifact common when mounting large tissue sections. Figure 5c shows similar field-expanded recoveries from each individual sensor, with green boxes indicating their physical dimensions and locations. Figures 5d-5f present a detailed comparison between MASI lensless raw data (Fig. 5d), single-sensor recovery (Fig. 5e), and MASI coherent synchronization of all sensors (Fig. 5f) for a region of interest in the brain section. The MASI coherent synchronization resolves the myelinated axon structure radiating outward from the central ventricle. This computational field expansion, combined with MASI's ability to operate without lenses at long working distances, enables a new paradigm for wide-field, high-resolution imaging that overcomes traditional optical system limitations[43].

The reported field expansion capability presents intriguing applications in data security and steganography. Since features outside the physical sensor area become visible only after proper computational padding and back-propagation, MASI creates a natural encryption mechanism for information hiding[44]. For example, a document could be designed where critical information -- authentication markers, security codes, or confidential data -- is positioned beyond the sensor's direct field of view. When capturing the raw intensity image, this information remains completely absent from the recorded data, creating an inherent security layer where the very existence of hidden content is concealed. This is demonstrated in Fig. 5, where Sensor 3's raw data and direct wavefield recovery give no indication that the computational reconstruction would reveal an entire brain section. Only an authorized party with knowledge of the correct wavefield recovery parameters, propagation distances, coded surface pattern, and padding specifications can computationally reconstruct and reveal this hidden content. This physics-based security approach offers advantages over conventional digital encryption by leaving no visible evidence that protected information exists in the first place.

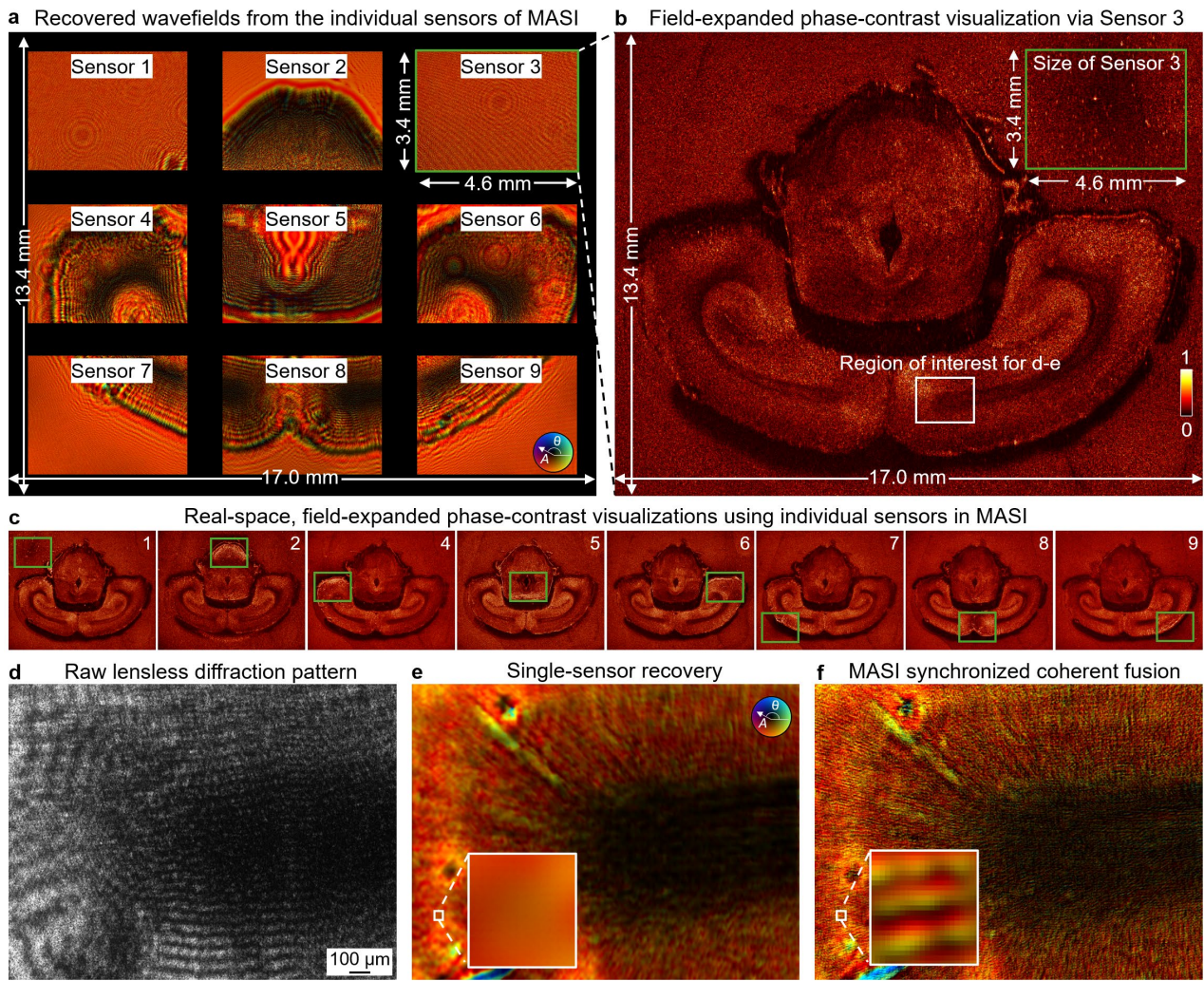

**Fig. 5| Computational field expansion for imaging a brain section. a**, Complex wavefields recovered at each of the coded sensors in a lensless transmission setup. Each individual sensor captures only a fraction of the object's diffracted field. The dark regions present the gaps between individual sensors. **b**, Real-space phase image of the brain section after propagation of Sensor 3's recovered wavefield to the object plane, showing that even a single sensor image expands to cover the entire brain section. **c**, Field-expanded recovered from individual sensors in MASI. Each recovered phase image highlights distinct aspects of the same sample. **d-f**, Comparative analysis of the region of interest (white box in b): raw lensless diffraction pattern (d), single-sensor recovery (e), and MASI coherent synchronization of all sensors (f). The coherent synchronization clearly resolves radiating myelinated axon bundles extending from the ventricle, while the single-sensor recovery shows limited resolution of these neural pathways, as highlighted in the insets.

**Computational 3D measurement and view synthesis**

Conventional lens-based 3D imaging typically relies on structured illumination techniques such as fringe projection, speckle pattern analysis, or multiple-angle acquisitions[45, 46, 47]. Similarly, perspective view synthesis often requires light field cameras with microlens arrays or multi-camera setups to capture different viewpoints[46, 47, 48]. Here we demonstrate that MASI enables lensless 3D imaging, shape measurement, and flexible perspective view synthesis through computational wavefield manipulation. Unlike the conventional lens-based approaches, MASI extracts three-dimensional information and generates multiple viewpoints through pure computational processing of the recovered complex wavefield.

The concept behind MASI's 3D imaging capability leverages the fact that a complex wavefield contains the complete optical information of a 3D scene. For 3D shape measurements, we digitally propagate the recovered wavefield to multiple axial planes throughout the volume of interest. At each lateral position, we evaluate a focus metric that quantifies local sharpness across all axial planes[49]. By identifying the axial position with maximum gradient value for each lateral point, we create a depth map where each pixel's value represents the axial coordinate of best focus. This approach effectively transforms wavefield information into precise height measurements, as objects at different heights naturally focus at different propagation distances. The resulting 3D map reveals microscale surface variations across the entire field of view.

For perspective view synthesis, MASI employs a different approach than conventional light field methods. We first propagate the reconstructed wavefield from the object plane at real space to the reciprocal space using Fourier transformation. In the reciprocal space, different angular components of light are spatially separated. By applying a filtering window to select specific angular components and then inverse-propagating back to real space, we synthesize images corresponding to different viewing angles. Shifting this filtering window effectively changes the observer's perspective, allowing virtual 'tilting' around the object for visualization.

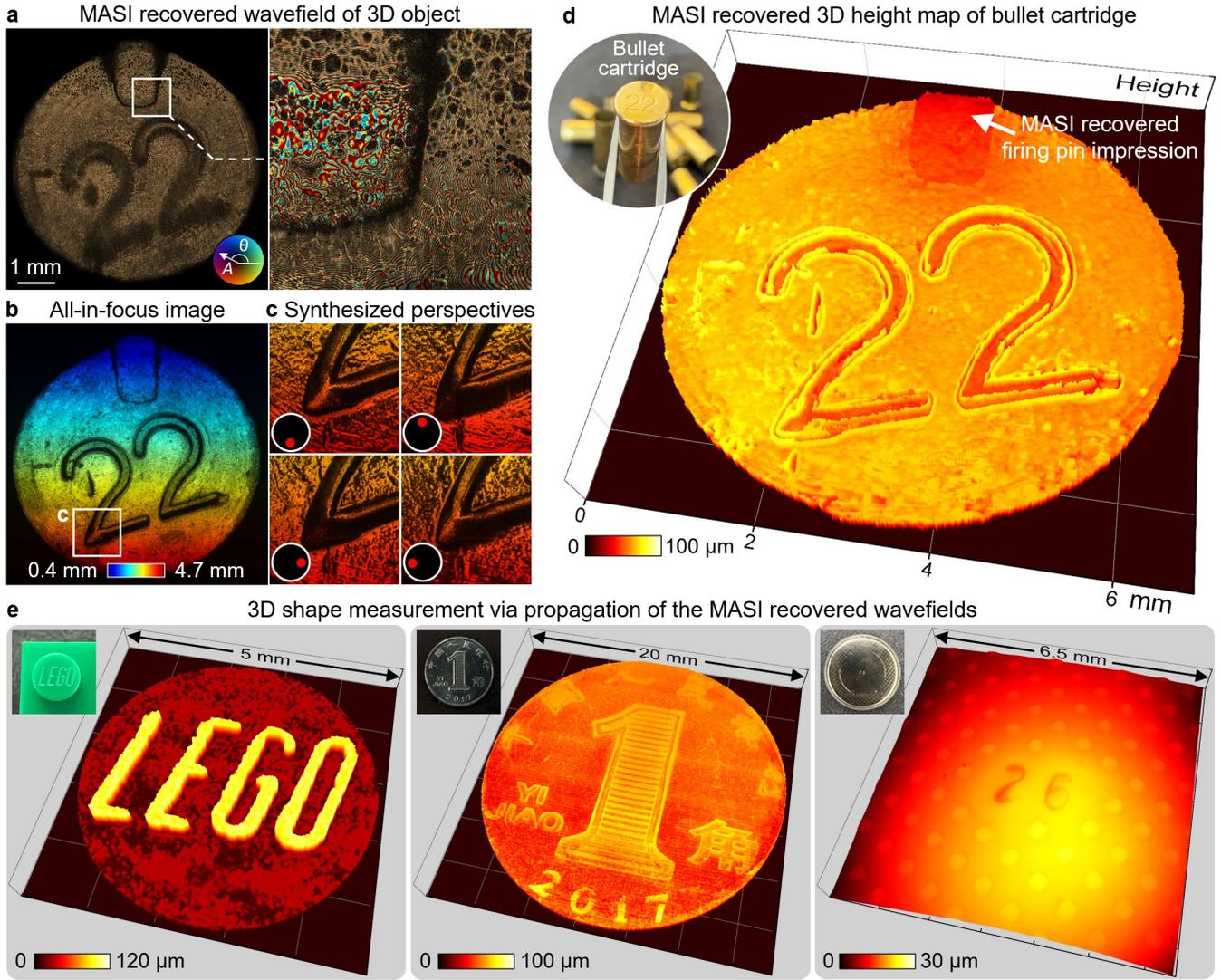

**Fig. 6| Computational 3D measurement and perspective view synthesis via lensless MASI. a**, MASI recovered wavefield of a bullet cartridge, with zoomed region showing detailed wavefield information. **b**, All-in-focus image with depth color-coding from 0.4 mm (blue) to 4.7 mm (red), providing comprehensive visualization of the cartridge's 3D structure. **c**, Synthetic perspective views generated by computationally filtering the wavefield at the reciprocal space, where different angular components of light are spatially separated. **d**, MASI recovered 3D map of the bullet cartridge, clearly revealing the firing pin impression and surface details for ballistic forensics. **e**, Demonstration of MASI's 3D measurement capabilities across objects with varying features. In each case, the 3D topography is reconstructed without mechanical scanning, highlighting MASI's potential for non-destructive testing and precision metrology. Supplementary Videos S3-S4 visualize the 3D focusing process. Supplementary Video S5 visualizes the generated different perspective views of the 3D object post-measurement.

Figure 6 demonstrates MASI's 3D measurement capabilities. In Fig. 6a, we show the recovered wavefield of a bullet cartridge using MASI. The recovered complex wavefield contains rich phase information that encodes the object's 3D topography. Figure 6b shows an all-in-focus image produced by our digital refocusing approach that combines information from multiple depth planes into a single visualization. This image provides both sharp features and depth information. Figure 6c shows different viewing angles generated by shifting the filtering window position in reciprocal space, revealing surface features that might be obscured from a single perspective. Figure 6d shows the recovered 3D height map revealing the firing pin impression and microscopic surface features, capturing critical ballistic evidence that could link a specific firearm to a cartridge casing. To demonstrate the versatility of MASI's 3D measurement capabilities, Figure 6e shows objects with varying feature heights and dimensions, including a LEGO brick with raised lettering, a coin with fine relief details, and a battery with subtle topography. The ability to generate 3D measurements and synthetic viewpoints through purely computational means represents an important advantage over conventional lens-based 3D imaging approaches.

## Discussion

We have introduced MASI, a lensless and scalable imaging architecture that transforms optical synthetic aperture imaging by computationally addressing the longstanding synchronization challenge. Unlike traditional systems that rely on precise interferometric setups, MASI employs distributed arrays of independently functioning coded sensors whose complex wavefields are computationally synchronized and fused entirely in real space. This multiscale approach allows MASI to surpass the diffraction limit of individual sensors without the stringent physical synchronization and overlapping measurement constraints that have historically limited practical deployment.

A key concept enabling MASI's capabilities is its computational phase synchronization strategy that fundamentally transforms how aperture synthesis can be implemented. While conventional synthetic aperture systems establish phase coherence through overlapping measurement regions[10, 37] or interferometry[50], MASI strategically optimizes only a single global phase offset per sensor. This design choice is critical as each individual sensor's ptychographic reconstruction recovers hundreds of megapixels of complex data, attempting pixel-level phase optimization would create an intractably large parameter space riddled with local minima. By restricting the degrees of freedom to just one parameter per sensor, MASI achieves a computationally efficient synchronization process that can readily scale to much larger arrays. By maximizing the integrated energy of the reconstructed wavefield, MASI ensures constructive interference among independently acquired wavefields -- analogous to wavefront shaping techniques in adaptive optics but implemented entirely in software. Consequently, MASI eliminates the need for reference waves or overlapping measurement areas, significantly simplifying implementation and enabling flexible, distributed sensor placement -- capabilities previously unattainable without complex interferometric configurations.

The scalable capability of MASI shares conceptual similarities with the Event Horizon Telescope (EHT), which synthesizes a large-scale aperture from separated radio receivers[40]. While the EHT maintains phase coherence through atomic clocks and precise timing synchronization, MASI replaces this hardware complexity with computational phase synchronization. In MASI, each sensor operates as an independent unit without requiring any overlap or shared measurement regions with neighbouring sensors. The architectural independence enables flexible sensor placement at various separations. This makes MASI particularly attractive for long-baseline optical imaging applications where maintaining interferometric coherence would be prohibitively challenging.

Like the sparse apertures employed in EHT, MASI faces the challenge of incomplete frequency coverage -- the physical gaps between sensors result in certain spatial frequencies not being recorded, analogous to the missing baselines in interferometric telescope arrays. Drawing inspirations from black hole imaging via EHT, MASI can potentially incorporate object-plane priors or constraints[40], regularization techniques[40], as well as neural field representation[35] to partially compensate for missing frequency information while preserving high-fidelity data from the measured regions.

Beyond addressing frequency gaps, MASI demonstrates inherent robustness to noise through its measurement strategy. The multiple sub-pixel shifted measurements acquired by each sensor provide overdetermined constraints that help suppress noise during iterative phase retrieval[37]. While our current implementation employs standard ptychographic reconstruction algorithms[41] without explicit regularization, the system's noise resilience could be further enhanced for low-light environments. Future implementations could incorporate feature-domain optimization[51, 52] and advanced regularization strategies such as total variation constraints for preserving edge features or sparsity priors in appropriate transform domains[53, 54], particularly beneficial for imaging under low-light conditions. These enhancements would suppress the noise and help to interpolate the missing frequency content problem discussed above.

The real-space coherent fusion approach of MASI also differs from traditional reciprocal-space aperture synthesis methods[4, 10]. By propagating the recovered wavefields back to the object plane and coherently fusing them, MASI naturally achieves two advantages. First, it eliminates the need for complicated reciprocal-space alignment or calibration, simplifying the overall imaging workflow. Second, this computational approach intrinsically encodes three-dimensional depth information, allowing MASI to achieve lensless 3D imaging and synthetic viewpoint generation without axial scanning.

MASI's scalability depends primarily on the imaging configuration. For far-field applications, the array size has minimal physical constraints, potentially enabling long-baseline configurations with widely separated sensors. However, for near-field imaging, practical limits arise from the angular response of image sensor pixels. Standard image sensors maintain high sensitivity (>0.9 relative response) for incident angles below less than 25°, but efficiency drops significantly beyond 45 degrees[55]. Using 60-degree as a practical limit, the minimum working distance $wd$ scales as $wd = d/(2 \cdot \tan(60°)) \approx 0.3d$, where $d$ is the array diameter. For example, a 10 cm array would require a minimum working distance of ~3 cm to maintain adequate signal at peripheral sensors. To mitigate these limitations, peripheral sensors can be tilted toward the object, similar to how LED arrays are tilted in Fourier ptychography to improve light delivery efficiency[56]. By redirecting each sensor's angular acceptance cone toward the object plane, we can maintain near-normal incidence even for sensors at extreme array positions. Additionally, the vignetting effect caused by angular response drop-off is predictable and can be computationally compensated during reconstruction. Since the pixel angular response characteristics can be well-characterized[55], the signal attenuation at each sensor position can be modelled and corrected based on the known incident angles.

MASI's demonstrated capability to resolve sub-micron features at ultralong working distances (~2 cm) can advance current optical imaging capabilities. Traditional optical microscopes typically require specialized objectives and short working distances (less than a millimeter) to achieve comparable resolutions. In contrast, MASI's lensless architecture maintains high spatial resolution even at centimeter-scale working distances, expanding its practical utility in fields such as industrial inspection, biological imaging, and forensic analysis, where sample accessibility and large-area imaging capabilities are critical.

Another achievement by MASI is the computational field expansion capability, as demonstrated with fingerprint and brain section imaging. By padding the reconstructed wavefields and back-propagating them computationally, MASI transforms limited sensor areas into much larger virtual imaging systems. This approach leverages the intrinsic reciprocal relationship between angular sensing and spatial reconstruction: angular information captured at the sensor plane naturally encodes spatial features beyond the sensor's immediate field-of-view. This computational field expansion capability also offers intriguing applications in data security and steganography. Because information outside the physical sensor area becomes visible only through computational reconstruction, MASI creates a natural physical-layer encryption mechanism. Hidden information embedded outside the direct sensor footprint remains entirely undetectable in the raw captured data, effectively concealing its existence. Only authorized parties possessing the correct computational parameters -- such as coded aperture patterns, padding extent, and propagation distances -- can reveal this concealed information. This inherent security feature could form the basis of novel verification methods for secure documents, anti-counterfeiting measures, and covert communications channels.

In MASI, coherent illumination is essential to produce high-contrast diffraction patterns that enable robust ptychographic reconstruction. In current implementation, laser diode is used to generate coherent illumination. For narrowband illumination with partial coherence (such as LEDs), reconstruction is indeed feasible -- LEDs have been successfully used as sources for both Fourier ptychography[10] and spatial-domain lensless ptychography[57].

These implementations typically employ mixed-state ptychographic algorithms[58] to account for partial coherence effects. For broadband illumination, while direct imaging would suffer from wavelength-dependent propagation and chromatic blur, narrow-band filters can be incorporated in the detection path to select specific wavelength ranges.

Future developments of MASI include extending the operational wavelength into infrared, terahertz, or X-ray regimes, which could open entirely new domains of remote sensing, security scanning, and advanced materials characterization. Incorporating polarization detection offers another dimension that would enable polarimetric imaging of birefringent or dichroic samples[59, 60], adding diagnostic power for biological tissues and advanced composites. MASI's architecture also extends to endoscopic applications, where multiple thin fiber bundles could function as distributed coded sensors[61, 62]. By computationally synthesizing wavefields from independently positioned fiber bundle tips, MASI could achieve super-resolution imaging through narrow channels. Scalability remains one of MASI's most compelling advantages -- its computational phase synchronization readily extends to larger sensor arrays with only linear increases in complexity, contrasting with the burdens faced by conventional interferometric systems. This feature could facilitate the construction of long-baseline optical arrays for remote sensing or aerial surveillance. Furthermore, the ability to place sensors at flexible separations and angles creates new possibilities for multi-perspective imaging, high-throughput industrial inspections, and planetary monitoring systems. The computational synchronization principle developed in MASI could also transform existing synthetic aperture radar and sonar systems, which currently rely on precise coherence maintenance over long baselines. By adopting MASI's post-measurement phase synchronization approach, these systems could potentially operate with relaxed hardware synchronization requirements, enabling more flexible deployment scenarios while maintaining or even enhancing imaging performance through computational means.

## Materials and methods

**MASI prototype**

The MASI prototype integrates an array of image sensors (AR1335, ON Semiconductor), each with 13 megapixels with 1.1-µm pixel size. To implement ptychography for wavefield recovery, we replace the coverglass of the image sensor with a coded surface with both intensity and phase modulation. These coded surfaces function as deterministic probe patterns analogous to those used in conventional ptychography, enabling complex wavefield recovery from intensity-only measurements[17, 37]. Two compact piezoelectric stages (SLC-1720, SmarAct, measured 22 mm × 17 mm × 8.5 mm) were adopted to provide controlled lateral movement of the MASI device. The range of movement is about 30 µm by 30 µm with 1-2 µm random varying step size. During operation, the MASI system captured images at 20 frames per second under continuous motion of the piezo stage, enabling efficient data acquisition without requiring positioning. Sample illumination is provided by a 10-mW laser diode, with exposure times in the millisecond range to minimize motion blur during the continuous acquisition process. In current implementations, we typically acquire 300 frames as a ptychogram dataset in ~15 seconds. The MASI's compact form factor and straightforward optical design facilitate both laboratory measurements and field deployment scenarios.

**Calibration experiment and real-space alignment**

Coded sensor calibration is important for MASI's coherent wavefield synthesis. Our calibration procedure consists of two major steps: coded surface characterization and sensor position determination. First, we characterized each sensor's coded surface pattern using blood smear samples as test objects. By acquiring and reconstructing diffraction patterns from these test objects under controlled conditions, we precisely mapped the amplitude and phase modulation characteristics of each sensor's coded surface[42]. With the coded surfaces fully characterized, we developed a robust one-time calibration procedure to determine the three-dimensional positioning parameters $(x_s, y_s, h_s)$ for each sensor in the array. This calibration employs point source objects that generate well-defined spherical wavefields. The calibration process follows three progressive refinement steps: 1) Initial axial distance estimation: We propagate each sensor's recovered wavefield through multiple planes and identify the position of maximum intensity concentration, providing a first approximation despite potential truncation artifacts. 2) Lateral position determination: We cross-correlate the back-propagated point-source reconstructions between each peripheral sensor and a designated reference sensor (typically the central sensor), establishing relative lateral displacements with sub-pixel precision[63]. 3) Parameter refinement through phase optimization: Leveraging the known properties of spherical wavefronts, we construct a phase-based optimization function that minimizes the difference between each recovered wavefield's unwrapped phase and the theoretical spherical wavefront model. This optimization significantly improves calibration accuracy by incorporating phase information and accommodating the three-dimensional nature of the sensor array. Detailed calibration procedures are provided in Supplementary Note 2.

**Computational phase synchronization**

A key innovation of MASI lies in its ability to synchronize the phase of independently recovered wavefields requiring no overlapping measurement regions. Conventional synthetic aperture systems typically depend on shared fields of view or interferometric measurements to establish phase relationships, posing major scalability challenges. In MASI, each sensor recovers only a local wavefield with an arbitrary global phase offset. We designate one sensor (e.g., the central sensor) as a reference and assign it zero phase offset. For the remaining sensors, we iteratively determine optimal offsets by maximizing the total intensity of the fused wavefield at the object plane:

$$\{\varphi_s\} = \underset{\varphi_s}{\mathrm{argmax}} \sum_{x,y} |O_{recover}(x,y)|^2 \tag{5}$$

This procedure is implemented by adjusting each sensor's phase offset sequentially while keeping the others fixed, and converges within a few iterations. By aligning wavefields to maximize constructive interference, MASI ensures robust phase synchronization purely through computation, circumventing the need for carefully matched optical paths. Supplementary Note 3 and Supplementary Video S1 detail the algorithm and show its convergence in practical experiments.

**Tilted-plane wavefield propagation**

In reflection-mode imaging, a tilted sensor plane introduces additional complexity because the measured wavefield no longer aligns with the sample surface. MASI addresses this by applying a tilt-plane wavefield correction in the Fourier domain, which transforms the measured tilted wavefield into an equivalent horizontal-plane representation. Suppose the spectrum of the recovered wavefield on the tilted sensor plane is $\widehat{W}_{tilt}(k'_x, k_y)$, where $k'_x$ is the frequency variable corresponding to the sensor's tilted $x$ axis. We can convert $\widehat{W}_{tilt}(k'_x, k_y)$ to the horizontal-plane spectrum $\widehat{W}(k_x, k_y)$ via

$$\widehat{W}(k_x, k_y) = \widehat{W}_{tilt}(k'_x \cos\theta - k'_z \sin\theta, k_y) \tag{6}$$

where $k'_z = (k^2 - k'^2_x - k^2_y)^{1/2}$ enforces the free-space dispersion relationship in the tilted coordinate system. Because this transformation modifies the coordinate scaling, an amplitude correction factor (Jacobian) is required for compensating for the altered sample density.:

$$J(k'_x, k_y) = \cos\theta + \frac{k'_x}{k'_z}\sin\theta \tag{7}$$

An inverse Fourier transform then yields the horizontal plane wavefield[64]:

$$W(x,y) = \mathcal{F}^{-1}\{\widehat{W}(k_x, k_y) \cdot |J(k'_x, k_y)|\} \tag{8}$$

This approach systematically compensates for the sensor's geometric tilt, reconstructing the reflection-mode wavefield as if captured by a non-tilted, horizontal sensor array. By incorporating these spectral transformations, MASI retains its lensless simplicity while supporting flexible sensor orientations in a wide range of reflection-based imaging scenarios. Supplementary Fig. S9 shows the tilted propagation method for the resolution target captured in reflection mode.

**3D measurement**

To generate accurate 3D maps from MASI measurements, we utilize the complex wavefield's ability to be digitally refocused to different depths. After the wavefield is recovered and coherently fused, we numerically propagate it across a range of axial positions to form a stack of amplitude images. At each lateral position in the field of view, we identify the optimal focus plane by maximizing an amplitude gradient metric[49], which peaks when local features are sharply focused. By recording the depth at which each pixel's focus metric is maximized, we obtain a continuous 3D surface profile of the sample. Using a standard axial resolution characterization methodology[65], we quantified MASI's depth discrimination capability to be approximately 6.5 μm, as demonstrated in Supplementary Fig. S13. This axial resolution enables precise 3D topographic mapping across various sample types and geometries. Additional 3D reconstructions showcasing this capability for different objects are presented in Supplementary Figs. S14-S16, with dynamic visualizations of the 3D focusing process available in Supplementary Videos S3-S4.

MASI's performance characteristics offer unique advantages compared to conventional 3D profilometry methods. While standard techniques such as interferometry and confocal microscopy can achieve high axial resolution, they typically require short working distances and have restricted field of view. MASI provides a compelling alternative by achieving micron-level axial resolution with centimeter-scale working distances and fields of view. Furthermore, the ability to computationally generate multiple viewing perspectives post-measurement, as shown in Fig. 6 and Supplementary Fig. S17, distinguishes MASI from conventional profilometry systems that typically provide only top-down height maps. This combination of long working distance, large field of view, and computational flexibility positions MASI as a versatile alternative for 3D metrology applications.

**Post-measurement perspective synthesis**
After reconstructing the high-resolution object wavefield $O_{recover}(x, y)$ by MASI, we synthesize different perspectives through numerical processing in the far-field domain. Specifically, the reconstructed complex wavefield is directly propagated to the far-field region by performing a Fourier transform operation. A rectangular pupil function $pupil(k_x, k_y)$ is then applied at different lateral positions to selectively sample specific angular components, effectively shifting the viewing angle. Inverse Fourier transforming the pupil-filtered wavefield yields $O_{view}(x, y)$, corresponding to each chosen viewpoint:

$$O_{view}(x, y) = \mathcal{F}^{-1}\{\mathcal{F}[O_{recover}(x, y)] \cdot pupil(k_x, k_y)\} \qquad (9)$$

By simply adjusting the pupil's position, one can generate diverse perspectives without additional data acquisition or hardware modifications. Supplementary Fig. S17 and Supplementary Video S5 illustrate these synthetic perspectives post-measurement.

**Acknowledgments**
This work was partially supported by the National Institute of Health R01-EB034744 (G. Z.), the UConn SPARK grant (G. Z.), National Science Foundation 2012140 (G. Z.), and Department of Energy SC0025582 (G. Z.). The content of the article does not necessarily reflect the position or policy of the US government, and no official endorsement should be inferred.

**Conflict of interest**
G. Z. is a named inventor of a related patent application. Other authors declare no competing interests.

**Author contributions**
G. Z. conceived the original concept of MASI and the procedures for computational phase synchronization. R. W. and Q. Z. developed the MASI prototypes, conducted the experiments, and analysed the data. R. W., Q. Z., Z. M. and Z. H. prepared the display items and Supplementary Videos. Q. Z. prepared the Supplementary Notes. M. M. and K. H. prepared the custom-designed flexible connection cables for the coded sensor array. D. B. and G. Z. wrote the manuscript with input from all authors. All authors participated in the discussion and interpretation of the results.

**Data availability**

The data supporting the findings of this study are available from the corresponding author on reasonable request.

**Code availability**

The computational phase compensation algorithm is described in detail in Methods and Supplementary Information. The source codes are available from the corresponding author on reasonable request.

**Supplementary information**

Supplementary Figs. S1-S17, Supplementary Notes 1-3, Supplementary Videos S1-S5.